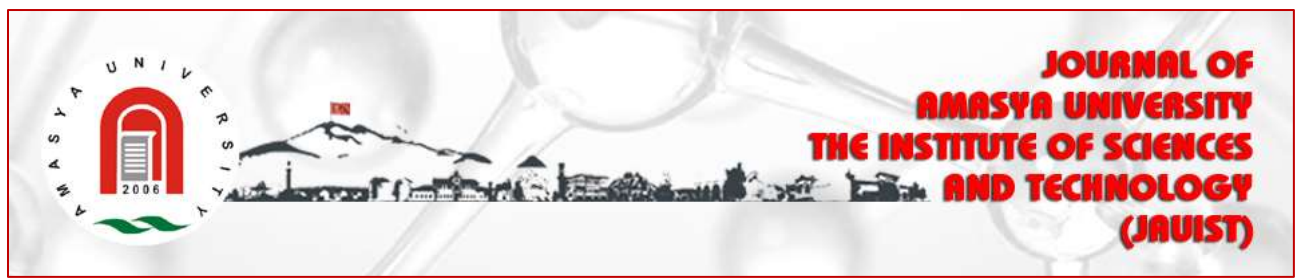

# BIBLIOMETRIC ANALYSIS OF THE WORLD SCIENTIFIC PRODUCTION IN CHEMICAL ENGINEERING DURING 2000-2011. PART 1: ANALYSIS OF TOTAL SCIENTIFIC PRODUCTION

Ruben MIRANDA (✉)[1], García CARPINTERO[2]

[1]Departamento de Ingeniería Química y de Materiales, Facultad de CC. Químicas, Universidad Complutense de Madrid. Avda. Complutense s/n, 28040 Madrid-Spain rmiranda@ucm.es
[2]Agencia de Evaluación de Tecnologías Sanitarias, Instituto de Salud Carlos III. Avda. Monforte de Lemos 5, 28029 Madrid-Spain eegarcia@isciii.es

**ABSTRACT**

A comprehensive bibliometric analysis of Chemical Engineering area has been carried out through the analysis of the scientific production covered in Web of Science during the period 2000-2011. Three complementary studies have been carried out. Part 1 analyzes total scientific production in the area. An important displacement of the scientific production to the Far East has occurred, mainly by the increase in publications from China (the world most productive country since 2008) but also from countries such as India and Iran. Although the share of publications from Europe, and especially from North America, have been decreased significantly, United States is still the country with the highest number of articles among the 1,000 most cited (31.5%), followed by Germany (8.4%) and China (7.5%). Switzerland, Italy, Netherlands, Singapore and Spain outstands as the countries with the highest number of cites per article and year and average impact factors of their publications. The international collaboration in the area is considerably high, especially within European countries (>40% publications are in international collaboration). The scientific production of the area is concentrated in a few journals (top 10 journals publishing around 30% publications and top 25 journals, around 50%) and publishers (Elsevier, Wiley-Blackwell, Taylor & Francis and the American Chemical Society). The countries with the highest number of institutions among the top 100 most productive are China and United States (12 each), followed by France (9), Japan (7) and United Kingdom (7). The top five institutions were from France, China, India, United States and Russia.

## ÖZET

2000-2011 döneminde Web of Science kapsamındaki çalışmalar üzerinden Kimya Mühendisliği alanının kapsamlı bir bibliyometrik analizi yapılmıştır. Üç tamamlayıcı çalışma yapılmıştır. Bölüm 1, bahsi geçen alandaki toplam bilimsel üretimi analiz etmektedir. Bilimsel üretimin Uzak Doğu'ya önemli bir yer değiştirmesi, esas olarak Çin'de (2008'den beri dünyanın en üretken ülkesi) ve ayrıca Hindistan ve İran gibi ülkelerde yayınların artmasıyla meydana gelmiştir. Avrupa'dan ve özellikle Kuzey Amerika'dan gelen yayınların payı önemli ölçüde azalmış olsa da, Amerika Birleşik Devletleri en çok atıf alan 1000 makale arasında (% 31,5) hala en çok makaleye sahip ülkedir, onu Almanya (% 8,4) ve Çin (% 7.5) izlemektedir. İsviçre, İtalya, Hollanda, Singapur ve İspanya, makale ve yıl başına en çok atıf yapan ülkeler ve yayınlarının ortalama etki faktörleri olarak öne çıkmaktadır. Bölgedeki uluslararası işbirliği, özellikle Avrupa ülkeleri içinde oldukça yüksektir (>% 40 yayınlar uluslararası işbirliği içindedir). Bölgenin bilimsel üretimi birkaç dergide (yaklaşık % 30 yayın yapan ilk 10 dergi ve yaklaşık % 50'si en iyi 25 dergi) ve yayıncılarda (Elsevier, Wiley-Blackwell, Taylor & Francis ve American Chemical Society) yoğunlaşmıştır. En verimli 100 kurum arasında en yüksek kurum sayısına sahip ülkeler Çin ve Amerika Birleşik Devletleri (her biri 12), ardından Fransa (9), Japonya (7) ve Birleşik Krallık (7) gelmektedir. İlk beş kurum Fransa, Çin, Hindistan, Amerika Birleşik Devletleri ve Rusya'dandır.

**Anahtar Kelimeler:** Kimya mühendisliği; bibliyometri; bilimsel üretim; atıflar; etki faktörü; uluslararası işbirliği; araştırma kuruluşları

## 1. INTRODUCTION

A possible definition of Chemical Engineering is that provided by Gillet (2001): "Chemical Engineering is the conception, development, design, improvement and application of processes and their products. This includes economic development, design, construction, operation, control and management of plant for these products together with research and education in these fields". Chemical Engineering synthesizes knowledge form several disciplines and interacts with researchers from multiple disciplines. However, Chemical Engineering has demonstrated a unique ability to synthesize diverse forms of knowledge from applied sciences and other engineering disciplines into cohesive and effective solutions to many societal needs (National Research Council, 2007). In this sense, Chemical Engineering covers a wide-ranging set of social interests and needs, including health, habitable environment, transportation, communications, agriculture, clothing and food, national defense and security, and various life amenities (National Research Council, 2007).

Chemical Engineering has enabled the science of Chemistry to achieve a high level of significance. According to the International Council of Chemical Associations, more than 20 million people around the globe are employed directly or indirectly by the chemical industry. In 2011, worldwide, it was estimated that world sales of chemicals amounted to over $3500 billion (International Council of Chemical Associations, 2014). In the United States, for example, over US $400 billion of products rely on innovations from chemical engineering and the business of chemistry supports nearly 25% of the US GDP (American Chemistry Council, 2014).

Although expensive and time-consuming, research and development is crucial to the chemical industry evolution. To keep competitive, the industry must find new products which enhance the quality of life adapt rapidly to changes in consumer demand around the world. At worldwide level, capital spending in research and development activities was around US$ 374 billion in 2011, with a capital intensity of 7.5% (ratio of capital spending to sales) (International Council of Chemical Associations, 2014). Furthermore, only in 2013 in United States, chemical companies invested $56 billion in R&D to support innovation and around 17% of United States patents are chemistry or chemistry-related (American Chemistry Council, 2014).

Bibliometry is a scientific discipline whose object is the analysis, classification and evaluation of the production and consumption of scientific information by quantitative and statistical methods. Bibliometry offers quantitative and objective information on the activity in science and technology areas, regarding their volume, evolution or visibility. Bibliometric analysis of the scientific activity is based on the assumption that realization of scientific activities and diffusion of the results are always joint activities. In this sense, the classic model input-output used to describe the scientific research process suggests that publications, as articles or monographs, can be used to represent the results of science (Russel and Rousseau, 2002). Therefore, scientific production, measured in terms of the number of publications, can be quantified and analyzed to determine the size and nature of the scientific activity measured at global, national or regional levels (macro level) or at institutional or research group levels (micro level). Bibliometric analysis is a valuable tool to understand the evolution and patterns followed by the science and to design new policies strategies to increase the international visibility of scientific activity. Bibliometry has been applied to a great variety of scientific disciplines. However, in the area of Chemical Engineering, there is a lack of comprehensive studies at worldwide level using a sufficient number of bibliometric indicators. Most of the previous studies focused only in a specific geographical region or country, e.g. the publications by Asian researchers (Yin, 2009; Chun et al., 2009), Chinese researchers (Fu et al., 2014), Spanish researchers (García-Carpintero and Miranda, 2013), Indian researchers (Modak and Madras, 2008), Mexican and other Latin American researchers (Rojas-Sola and San-Antonio-Gómez, 2010a), Uruguayan and other Latin American researchers (Rojas-Sola and San-Antonio-Gómez, 2010b), or researchers affiliated to a single institution such as the Institute of Chemical Engineering in Taiwan (Chang and Cheng, 2012). Furthermore, there is also literature in which the bibliometric analysis focused on the scientific output in specific journals such as the study of Schubert (1998). There are also studies focused only in specific bibliometric indicators, especially those related to the top-cited or high-impact papers published in the Chemical Engineering area, i.e. the works carried out by Ho (2012) and Chuang et al. (2013). However, there is a lack of discussion comparing the results obtained through the analysis of the total scientific output in this area and the most cited papers.

The objective of the present study is to improve the knowledge in these two areas together with a thorough analysis of the research areas and trends in Chemical Engineering. First, Part 1 of this study will analyze the world scientific production in Chemical Engineering from 2000 to 2011, distinguishing between different regions and countries and the evolution of the main indicators with time. Especifically, it will be analyzed: a) the number of publications, b) quality indicators of the publications based on the number of citations received (impact factor, received citations and the most cited publications in the area), c) the degree of international collaboration, d) the journals with the highest scientific production, and e) the most productive organizations. On the other side, Part 2 of this study will focus on the bibliometric analysis of the 1,000 most cited papers in the Chemical Engineering area from 2000 to 2011 and their results will be compared together with the results obtained in Part 1. Finally, Part 3 of this study will analyze the most important thematic areas and research trends through the analysis of the article title words, author keywords, keyword plus®, both on the total scientific production and the 1,000 most cited papers in the period 2000-2011.

## 2. METHODOLOGY

Scientific production in the area of Chemical Engineering has been analyzed through the use of the Web of Science® Core Collection database. Web of Science (WoS) is an on-line service for scientific information, developed by the Institute for Scientific Information (ISI), and now integrated in Clarivate Analytics. WoS facilitates the access to a set of bibliographic databases from different areas, which includes publications on Science (Science Citation Index), Social Sciences (Social Science Citation Index) and Art and Humanities (Arts and Humanities Citation Index). Besides, WoS publishes the Journal of Citation Report (JCR), one of the most prestigious databases for cites analysis of the publications.

Present study has been circumscribed to the journals indexed in the area of "Engineering, Chemical" of the WoS, the number of these journals varying from 110 to 135 during the analyzed period (2000-2011). The analyzed publications include all the papers which have been published in the journals which were indexed in the Chemical Engineering area according to JCR 2011, although these journals were not previously indexed in this area. In such way, data along the analyzed period can be compared without being influenced by the number of journals indexed in this area with time. In addition, only articles and reviews document types were considered.

A total of 213,264 documents were analyzed: 228,065 articles (98.5% total publications) and 3,199 reviews (1.5% total publications). There were used 16 different languages in these documents, however, most of the publications were in English (199,782 publications, 93.68%), German (5,318, 2.49% total publications) and Polish (2,872, 1.35%). These languages were followed by Chinese (0.89%), Romanian (0.63%), Japanese (0.63%), Spanish (0.13%), Serbo-Croatian (0.11%) and French (0.07%), the other languages used each in less than 50 publications in total (< 0.02% total publications).

Although the journals indexed in the thematic area of "Engineering, Chemical" do not represent 100% of the scientific production in Chemical Engineering, they represent a good sample of the research carried. In these sense, most of the previous works followed this approach (Rojas-Sola and San Antonio Gómez, 2010a, 2010b; Ho, 2012; Chuang et al., 2013; García-Carpintero and Miranda, 2013; Fu et al., 2014). Furthermore, JCR allows one journal to be indexed in several thematic areas, which facilitates some thematic areas related to the main thematic area which is analyzed to be taken into account. According to the data from JCR 2011, from the 133 total number of journals included, only 34 journals (25.6%) were indexed only in the Chemical Engineering area, while other 49 journals (36.8%) were indexed in two thematic categories, 38 journals (28.6%) in three thematic areas and finally, 9 journals (6.8%) and 3 journals (2.3%) were indexed in four and five thematic categories, respectively. These other JCR defined thematic categories, by order of importance, are: "Energy & Fuels", "Chemistry, Applied", "Chemistry, Multidisciplinary",

"Chemistry, Physical", "Materials Science, Multidisciplinary", "Engineering, Petroleum", "Engineering, Petroleum", "Polymer Science" and "Thermodynamics".

When the share of publications by nationalities is studied, it is important to take into account that the percentages are based on attributing one full point to each country in the case of internationally coauthored papers (Leydesdorff and Wagner, 2009). For this reason, world shares may add up to more than 100%. In the case of total scientific production by geographical areas, the data were adjusted to take this into account by removing duplicate papers with multiple national authorship (D. King, 2004). Besides, articles originating from England, Scotland, North Ireland and Wales were reclassified as being from the United Kingdom.

## 3. RESULTS AND DISCUSSION

### 3.1. Scientific production

World scientific production in the area of Chemical Engineering was 213,264 publications during the period 2000-2011. The annual production continuously increased from the year 2000 (13,797 publications) to 2011 (24,497 publications), at an average 7.1% annual growth (Fig. 1).

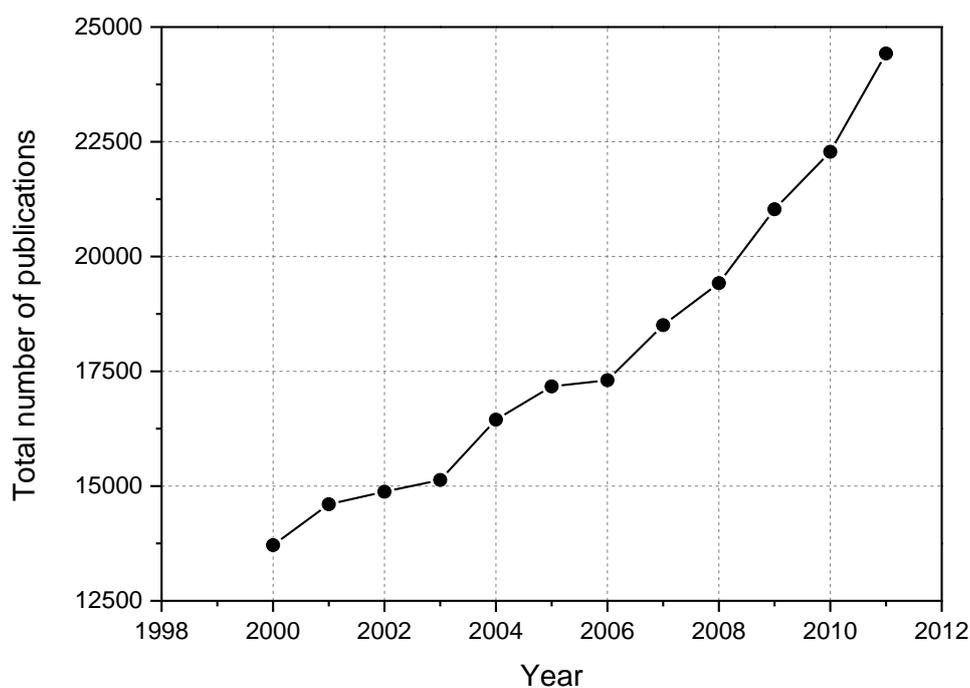

**Figure 1.** Evolution of total number of publications in the area of Chemical Engineering from 2000 to 2011.

### 3.1.1. Scientific production by regions

First, the scientific production by geographical areas was reviewed and the evolution of the share of these regions in the total scientific production with time is shown in Fig. 2.

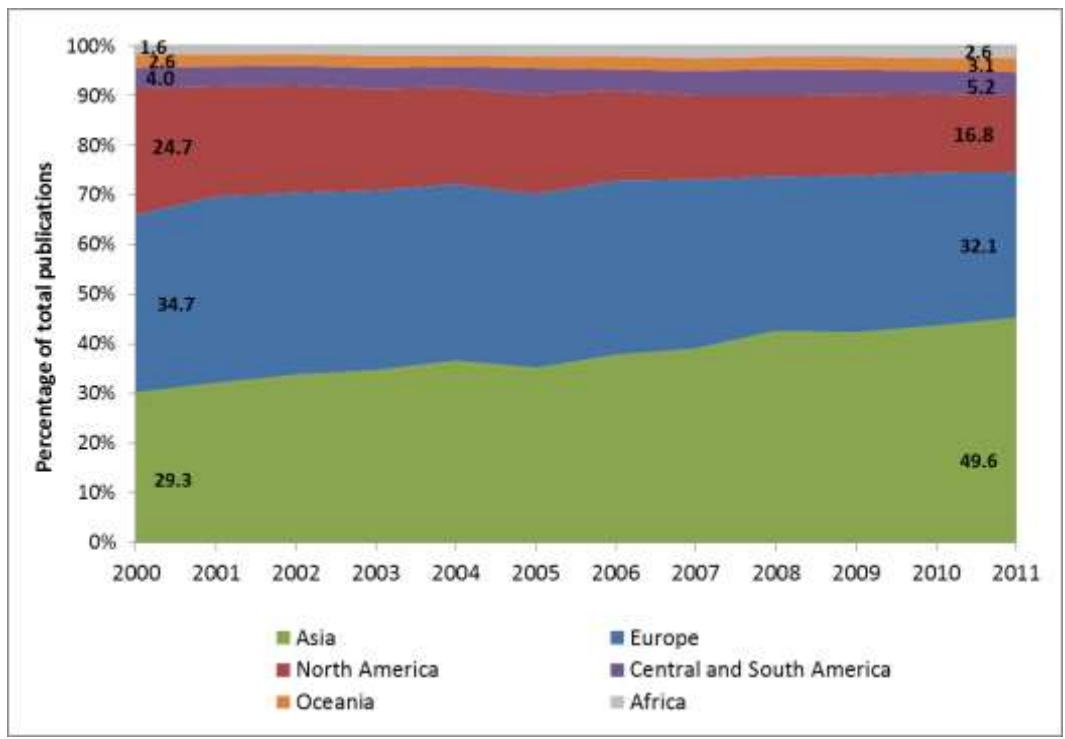

**Figure 2.** Evolution of the share of publications by geographical regions during 2000-2011.

North American scientific production (United States and Canada) was 40,569 publications (19.0% of world scientific production). A continuous increase in the number of publications was observed from 2000 (2,992 publications) to 2011 (4,113), however, the annual average growth was the lowest of the analyzed regions (3.4%). As a consequence, the share of North America in total publications decreased largely since 25.1% in 2000 to 17.1% in 2011.

Central and South American scientific production in 2000-2011 was 10,120 publications (4.7% total publications). It increased since 2000 (558 publications) to 2011 (1,285 publications) at an annual average growth rate of 11.8%. The importance of Central and South American share in total publications increased from 4.0% in 2000 to 5.2% total publications in 2011. The countries with the highest scientific production were Brazil, Mexico, Argentina and Chile, those four countries participating in more than 90% of the total production of the region.

European scientific production during the analyzed period was 73,730 publications. It increased from 4,791 publications in 2000 to 7,869 publications in 2011, which means an annual average growth rate of 5.8%. If Europe is considered as a whole, it represented a 34.6% of worldwide production during the period 2000-2011, well above countries such as United States (15.3%), China (12.2%) or Japan (6.0%). However, the importance of Europe in world scientific production has decreased slightly from 34.7% in 2000 (reaching a maximum of 36.5% in 2006) to 32.1% in 2011. The European countries with the highest importance in terms of scientific production were Germany, United Kingdom, France, Spain, Poland, Italy, the Netherlands and Romania, these eight countries participating in 83.5% of the publications with European authors.

Asian scientific production during the analyzed period was 86,404 publications. The number of Asian publications increased largely since 2000 (4,091 publications) to 2011 (12,250 publications) (annual average growth rate of 18.1%). In the analyzed period, Asia represented 40.5% of worldwide production, but its share in total publications have largely increased from 29.6% in 2000 to 50.0% in 2011. The Asian countries with the highest scientific production are China, Japan, India, South Korea, Turkey, Russia, Taiwan and Iran, these eight countries participating in 91.1% of the publications with authors from Asian countries.

African scientific production during the analyzed period was 4,492 publications. The number of publications increased since 2000 (218 publications) to 2011 (635), which means an annual average growth rate of 17.4%. In the analyzed period, the share of African scientific production in total production increasing from 1.6% in 2000 to 2.6% in 2011 (2.1% average during 2000-2011). The most productive African countries are Egypt, South Africa, Algeria, Tunisia, Nigeria and Morocco. All articles published by African countries have at least one author from any of these six countries.

Finally, the Oceanian scientific production during the analyzed period was 5,696 publications. The number of publications increased since 2000 (359 publications) to 2011 (771), which turned on annual average growth rate of 10.4%. In the analyzed period, Oceania represented 2.7% of worldwide production, the share of the Oceanian scientific production in total production being increased from 2.6% in 2000 to 3.1% in 2011. The scientific production by Oceanian countries is practically that from Australia, and in a lower extent, from New Zealand.

*3.1.2. Scientific production by countries*

Table 1 shows the total number of publications of the 30 most productive countries in the period 2000-2011.

Table 1. Scientific production of the 30 most productive countries during 2000-2011.

| Rank | Country | P2000-2011 | % world. | P2000 (rank in 2000) | P2011 (rank in 2011) | Average annual growth (%) |
|---|---|---|---|---|---|---|
| 1 | United States | 32,597 | 15.28 | 2,913 (1) | 3,248 (2) | 1.1 |
| 2 | China | 26,110 | 12.24 | 555 (6) | 4,714 (1) | 68.1 |
| 3 | Germany | 13,423 | 6.29 | 1,087 (2) | 1,232 (4) | 1.2 |
| 4 | Japan | 12,780 | 5.99 | 1,035 (3) | 1,098 (6) | 0.6 |
| 5 | United Kingdom | 10,611 | 4.98 | 894 (4) | 1,008 (10) | 1.2 |
| 6 | India | 10,145 | 4.76 | 463 (10) | 1,414 (3) | 18.7 |
| 7 | France | 9,964 | 4.67 | 656 (5) | 1,047 (8) | 5.4 |
| 8 | South Korea | 9,128 | 4.28 | 499 (8) | 1,034 (9) | 9.8 |
| 9 | Spain | 8,725 | 4.09 | 427 (11) | 1,170 (5) | 15.8 |
| 10 | Canada | 8,600 | 4.03 | 547 (7) | 939 (11) | 6.5 |
| 11 | Poland | 6,569 | 3.08 | 278 (15) | 724 (13) | 14.6 |
| 12 | Turkey | 5,937 | 2.78 | 239 (17) | 721 (12) | 18.3 |
| 13 | Russia | 5,460 | 2.56 | 480 (9) | 528 (18) | 0.9 |
| 14 | Australia | 5,122 | 2.40 | 328 (12) | 715 (14) | 10.7 |
| 15 | Italy | 4,922 | 2.31 | 327 (13) | 553 (17) | 6.3 |
| 16 | Taiwan | 4,825 | 2.26 | 310 (14) | 562 (15) | 7.4 |
| 17 | Iran | 4,403 | 2.06 | 85 (29) | 1053 (7) | 103.5 |
| 18 | Brazil | 4,317 | 2.02 | 275 (16) | 554 (16) | 9.2 |
| 19 | Netherlands | 3,936 | 1.85 | 240 (18) | 351 (20) | 4.2 |
| 20 | Romania | 3,400 | 1.59 | 188 (19) | 320 (22) | 6.4 |
| 21 | Portugal | 2,405 | 1.13 | 106 (24) | 317 (23) | 18.1 |
| 22 | Mexico | 2,262 | 1.06 | 88 (27) | 337 (21) | 25.7 |
| 23 | Singapore | 2,108 | 0.99 | 106 (25) | 253 (25) | 12.6 |
| 24 | Greece | 2,049 | 0.96 | 108 (23) | 202 (32) | 7.9 |
| 25 | Switzerland | 1,996 | 0.94 | 122 (22) | 211 (31) | 6.6 |
| 26 | Argentina | 1,918 | 0.90 | 123 (21) | 220 (27) | 7.2 |
| 27 | Sweden | 1,916 | 0.90 | 126 (20) | 217 (29) | 6.6 |
| 28 | Belgium | 1,685 | 0.79 | 91 (26) | 220 (28) | 12.9 |
| 29 | Finland | 1,489 | 0.70 | 74 (33) | 213 (37) | 14.0 |
| 30 | Egypt | 1,472 | 0.69 | 84 (31) | 158 (30) | 10.3 |
|  | Top 30 | 210,278 | 98.60 | 12,854 | 25,333 | 8.8 |
|  | World | 213,264 | 100.00 | 13,797 | 24,497 | 7.1 |

Note: P2000-2011 = Total publications 2000-2011; P2000 = Publications in 2000; P2011 = Publications in 2011.

Furthermore, the number of publications and rank of each country in 2000 and 2011, and the average growth rate in the number of publications were included. At least one of these 30 most productive countries was present in 98.60% of total publications in the area.

At worldwide level, the countries with the highest scientific production during the period 2000-2011 were, in this order, United States, China, Germany, Japan, United Kingdom, India, France, South Korea, Spain and Canada. However, there are very important differences between the beginning and the end of the analyzed period. In 2000, United States (2,913 publications), Germany (1,087), Japan (1,035), United Kingdom (894) and France (656) were the five countries with the highest number of publications in the area, representing together almost half (47.7%) of the total worldwide scientific production. Again in 2011, the five countries with the highest number of publications represented almost half (48.1%) of total publications. However, the countries and their relative rank of the top five most productive countries were quite different: China (4,714 publications), United States (3,248), India (1,414), Germany (1,232) and Spain (1,170).

Fig. 3 shows the evolution of the scientific production of the ten most productive countries. China became the most productive country in 2008, and already in 2011 had 45% more scientific publications than United States. The most recent data (publications in 2015), indicates this trend has not slowed down. In 2015, the number of publications of China was more than twice those of the United States: 8,530 vs. 3844 publications. In the analyzed period, scientific production of China increased from only 555 publications in 2000 to 4,714 in 2011, which means an annual average growth rate of 68.1%.

The shift from dominance from United States to China within global publication and citation shares have attracted many studies (Zhou and Leydesdorff, 2006; Leydesdorff and Wagner, 2009; Shelton and Foland, 2009, Leydesdorff, 2012). Considering the total number of publications in all scientific areas, a 2011 report from the Royal Society of United Kingdom reported that China was on a trend line to overtake the United States in terms of publications in 2013 (Clarke and Plume, 2011), however, Leydesdorff et al. (2014), suggested a date beyond 2020 for the cross-over as the exponential growth of Chinese scientific publications had slowed to linear growth rates during 2000s. By 2006, China was already the second largest nation in terms of publication within SCI-expanded, however, the United States was still by far the strongest nation in terms of scientific performance, especially in terms of highly cited papers and citation/publication ratios (Leydesdorff and Wagner, 2009).

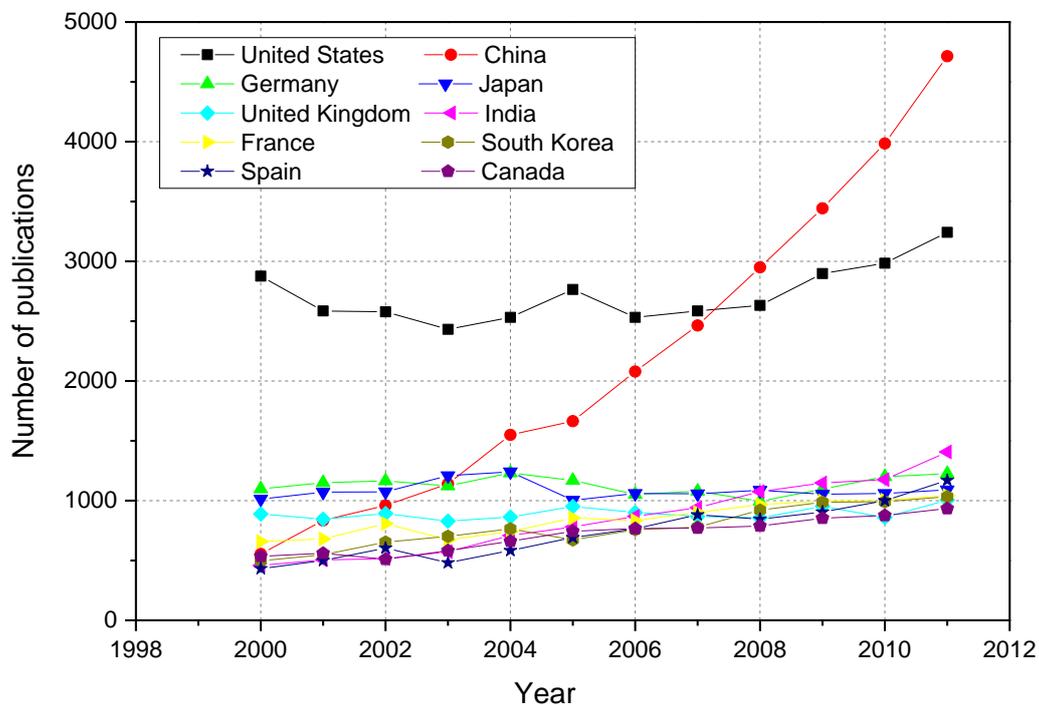

**Figure 3**. Evolution of the number of number of publications by year of the top 10 most productive countries during the period 2000-2011.

However, depending on the scientific database considered, the predominance of United States is less clear. For example, Kostoff (2008), with results from 2007, found that in almost all technical areas the United States had the lead in number of publications when SCI database was used. However, in INSPEC database, China attained parity with, or exceeded, with the United States already, and in Ei Compendex, China had even higher scientific production compared to the United States. As argued by Kostoff (2008), the reason is that SCI contains substantial biomedical research papers, where United States has a substantial intrinsic advantage, which are not included in more technical INSPEC and Ei Compendex.

Whatever the case, in some areas as in Chemical Engineering, the overhelming predominance of Chinese growth is clear. The crossover between the total number of publications of China and United States already happened before in 2008. Even more important, only six years after this crossover (2015), the number of publication is already more than double of the scientific production of the United States. This means the exponential growth of publications in this area still had not slowed down.

As commented before, a clear displacement of the scientific production in Chemical Engineering to the East was observed, mainly to China, but also to countries such as India and Iran. In the case of India, the scientific production increased from 463 publications in 2000, when India was in tenth position in terms of scientific production, to 1,414 publications in 2011 (annual average growth of 18.7%), and occupying in 2011 the third position in terms of scientific production. Iran is the country with the highest increase in the number of publications during the analyzed period, from 85 publications in 2000 to 1,053 publications in 2011 (average annual growth rate greater than 100%). In 2000, Iran occupied the $29^{th}$ position in terms of scientific production and in 2011 is already the $7^{th}$ country. The increase in publications from Iran is still growing exponentially, the most recent data from 2015 indicates Iran is already the $4^{th}$ country in terms of scientific production (1,751 publications), only behind China, the United States and India.

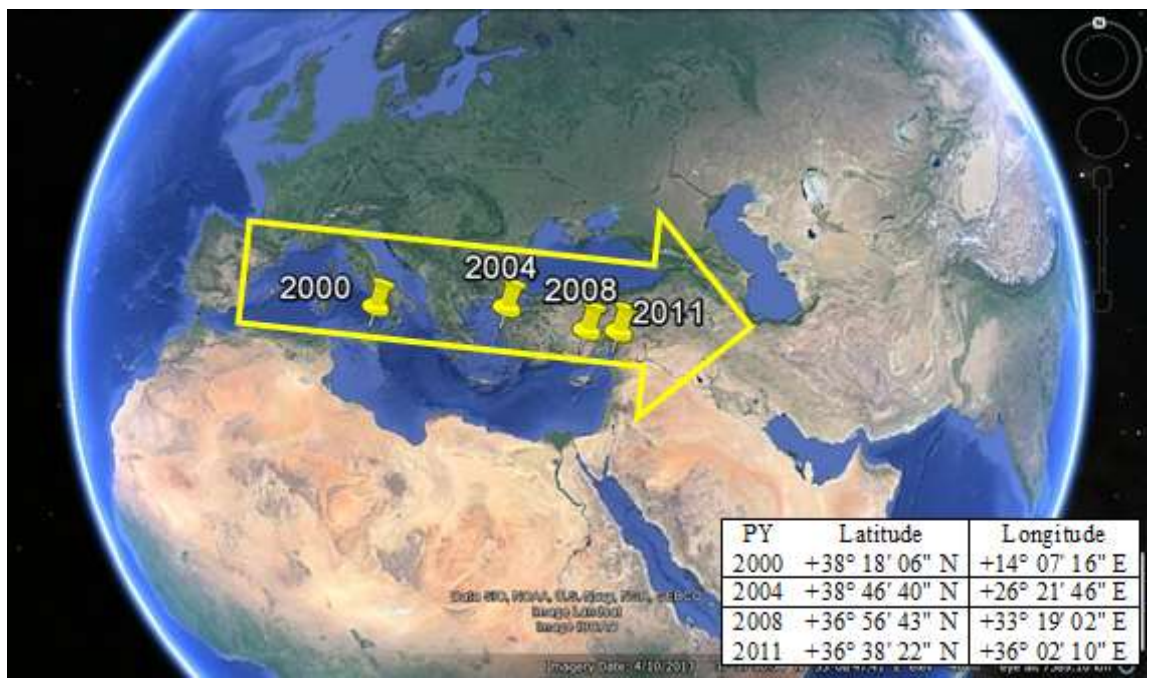

**Figure 4.** Gravity center of the worldwide scientific production of the 30 countries with the highest scientific production during the period 2000-2011. Source: Google Earth.

If the number of publications of the top 30 most productive countries during the period 2000-2011 is considered together with their geographical coordinates, a plot representing the "gravity center" of the scientific production in Chemical Engineering can be generated and its evolution with time can be analyzed (Fig. 4). The latitude and altitude of this "gravity center" have been calculated for the different years as follows:

$$(\text{Latitude}, \text{Altitude}) = \left( \frac{\sum_{i=1}^{j} N_i \cdot \text{Latitude}_i}{\sum_{i=1}^{j} N_i}, \frac{\sum_{i=1}^{j} N_i \cdot \text{Longitude}_i}{\sum_{i=1}^{j} N_i} \right)$$

where $N_i$ is the number of publications of each country i, and $\text{Latitude}_i$ and $\text{Longitude}_i$ are the geographical coordinates of the capital of each country.

### 3.2. Quality indicators of the scientific production

The number of citations and related indicators, such as impact factor, do not necessarily indicate the quality of a paper, however, it is a common measure of its impact and/or visibility in the area (Agarwall and Agarwall, 2007; Bornmann et al., 2012). It provides a marker of its recognition within the scientific community and frequently, the best manuscript can be considered the one most cited in peer-reviewed journals (Ho, 2012).

#### 3.2.1. Impact factor

The impact factor is a quality indicator for journals which is calculated from the number of cites received by papers published in the last two years compared to the number of articles published the present year. Although this is not a quality indicator for single papers but the journal in which it is published, it is commonly used to measure the quality of the papers despite articles published the same year and in the same journal can have very different citation rates.

Fig. 5 shows the average impact factor of the 30 most productive countries together with the world average and the top 30 most productive countries average in the period 2000-2011. It is important to take into account that the average impact factor of the publications in the area have increased largely with time, as occurred in other scientific areas, due to the increased number of documents covered by the database. In Chemical Engineering, the world average impact factor varied from 0.71 in 2000 to 2.51 in 2011 (average 1.37) and for the 30 most productive countries, from 0.79 in 2000 to 2.44 in 2011 (average 1.44). For this reason, the average impact factor for each country was calculated based on yearly averages and not based on the total number of publications. This avoids countries having a large share of articles in the last years of the period showing comparatively higher average impact factors than others as the impact factor of the journals increased largely during this period.

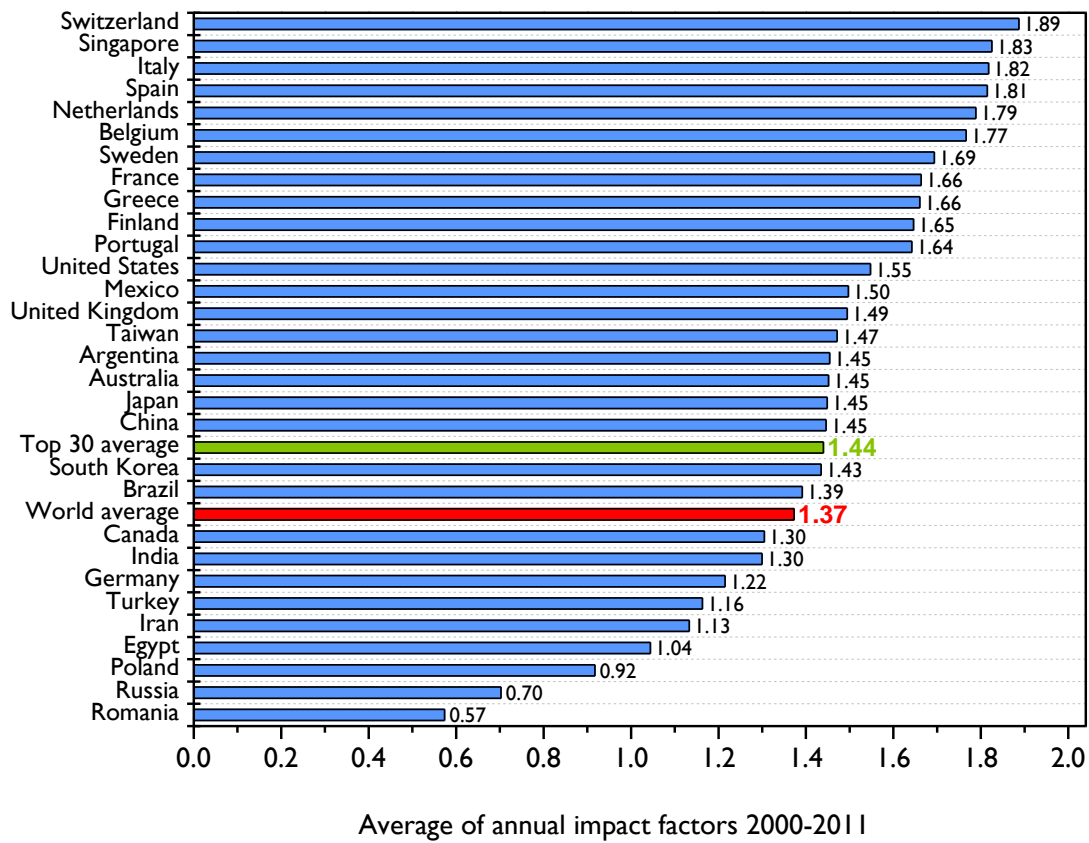

**Figure 5.** Average of annual impact factors of the publications during the period 2000-2011 of the 30 most productive countries in the area.

The average impact factor of the 30 most productive countries varied largely from 0.57 (Romania) to 1.89 (Switzerland). The countries with the highest average impact factor in their publications (> 1.7) are Switzerland (1.89), Singapore (1.83), Italy (1.82), Spain (1.81), Netherlands (1.79) and Belgium (1.77). On the other side, the countries with the lowest average impact factor on their publications (< 1.1) are Romania (0.57), Russia (0.70), Poland (0.92), Egypt (1.04), Iran (1.13) and Turkey (1.16). It is interesting notice that the average impact factors of the most productive countries were not the highest, in fact, the impact factor of the most productive countries such as United States, China, Germany, Japan, United Kingdom or India were similar and sometimes even slightly lower than the average impact factor for the top 30 most productive countries, varying from 1.22 to 1.55.

It is remarkable the existence of some small countries with low scientific production but with a high quality. Most of them are located in Europe, i.e. Switzerland, Netherlands, Belgium, Sweden, Finland or Denmark. Previous studies have recognized the important role played in the European science by these smaller northern countries, especially its contribution in the most cited papers. For instance, King (2004) demonstrated the six countries previously commented had a similar share in the most cited papers of the SCI database in the period 1997-2001 (12.7%) than United Kingdom (12.8%) or Germany (10.4%) although the total GDP of these countries is significantly lower. Already in the study of Glänzell and Schubert (2001), Switzerland, Netherlands, Sweden and Denmark were among the top five countries in terms of citations per publication. Similarly to the results presented in this study, Glänzell and Schubert (2001) also shown that countries such as Russia and Egypt were those with the lowest ratio citations per publication, which demonstrates this behavior is not particular of the area of Chemical Engineering but is general in the scientific production of these countries.

Traditionally, the quality of publications from China, measured by impact factor of the journals where published or by the number of citations received, has been lower than the world average and that of the traditional developed countries such as United States, Australia, France or United Kingdom, independently of the scientific area considered (Yi and Xi, 2008; Yi and Jie, 2011; Fu et al., 2013; Tan et al., 2014). However, last results indicates the increase in quality of Chinese publications, appearing to be competitive with countries such as France, Italy, Japan or Australia in terms of citations (Kostoff, 2008). In the present study, the impact factor of publications with Chinese authors is already about the same than the average of the top 30 most productive countries in the area of Chemical Engineering (1.45 vs. 1.44), with similar results to traditional countries such as United Kingdom, Japan or Australia.

### 3.2.2. Number of received citations

During the period 2000-2011, the 213,264 articles published in the area received a total of 1,970,755 cites since publication to August 2012, when the bibliographic data were downloaded from WoS. This means an average of 9.2 cites per article, the average number of citations decreasing continuously from 13.1 (articles published in 2000) to 1.4 (articles published in 2011). Obviously, the articles need time to be cited, thus the articles published earlier were at an advantage to gain more citations compared to those published at a later time (Lefaivre et al., 2011). To consider that, the average number of citations received per article and year since they were published was calculated. In this case, this value increased from 1.09 in 2000 to 1.86 in 2009 and then, decreased in 2010 and 2011 (average in 2000-2011, 1.54). The reason for this growth is the increase in the number of publications and journals covered in the area, also occurring in all the areas WoS, which increases largely the number of possible citing sources. In the case of the 30 most productive countries, the average cites received by each article and year has continuously increased from 1.34 (articles published in 2000) to 2.58 (articles published in 2012), being 1.98 cites the average for articles published during the period 2000-2011.

Fig. 6 shows the average citations per article and year since publication to 2012. The countries with the highest number of citations (>2.5) are: Switzerland (3.04), Singapore (2.95), Greece (2.75), Spain (2.65), Netherlands (2.65), Portugal (2.58) and Italy (2.57). On the other hand, the countries with the lowest citations (<1.6) are: Romania (0.64), Russia (0.66), Poland (0.93), Egypt (1.17), Iran (1.48), Mexico (1.53) and Argentina (1.57).

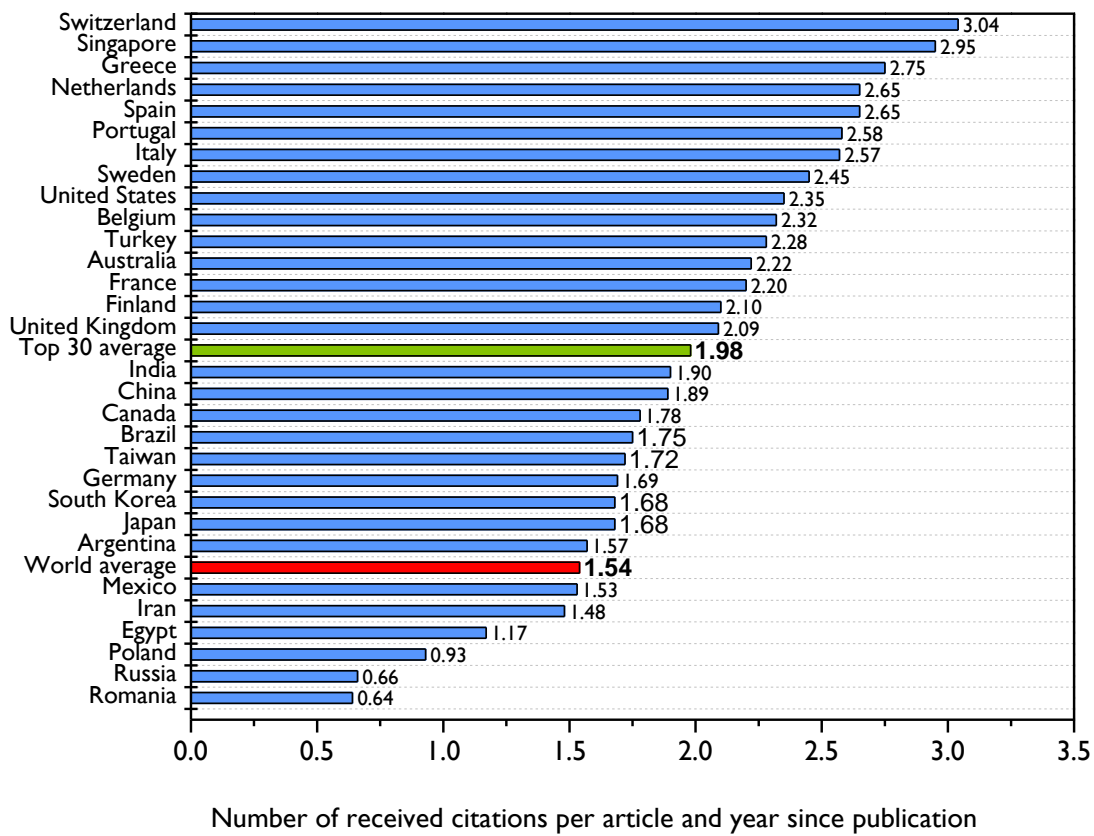

Number of received citations per article and year since publication

**Figure 6.** Total number of cites per article and per year since publication to 2012 of the 30 most productive countries.

Of the 10 most productive countries, only Spain (2.65), United States (2.35), France (2.20) and United Kingdom (2.09) had a greater value than the world average (1.98). However, there are many countries with low scientific production but having a number of citations well above the average, e.g. Switzerland (3.04), Singapore (2.95), Greece (2.75), Portugal (2.58) or Sweden (2.45). China performance (1.89) was better than the world average and similar to that of the top 30 most productive countries, which demonstrates the quality of Chinese publications is improving significantly in the last decade, not only in this area, as also observed in other studies (Kostoff, 2008; Leydesdorff and Wagner, 2009; Fu and Ho, 2013).

### 3.2.3. Most cited publication

The bibliometric analysis of the most cited papers in the area will be analyzed in detail in Part 2 of this study. In Part 1, only the total number of publications among the 1,000 most cited articles by nationalities were analyzed (Fig. 7).

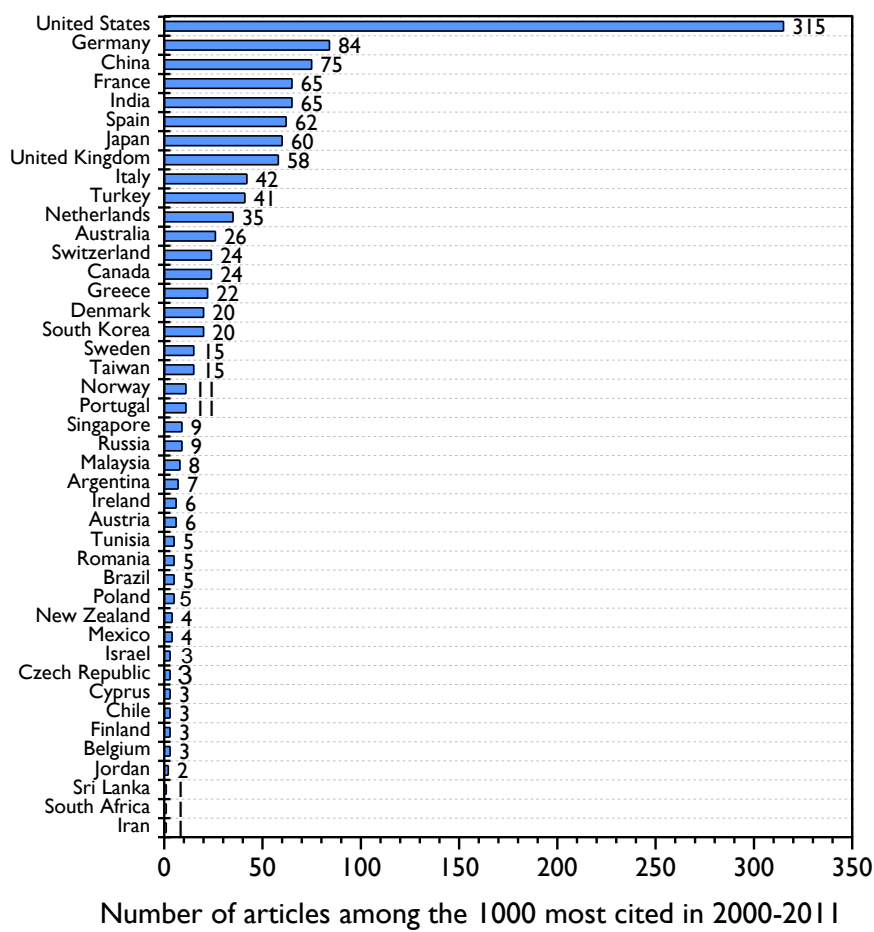

**Figure 7.** Number of articles among the 1,000 most cited in the area by countries during the period 2000-2011.

In this case, there is a large difference between the United States and the rest of the countries. A 31.5% of the 1,000 most cited papers have an author from the United States. The second and third country with more articles among the most cited were Germany and China, participating in 8.4% and 7.5% of total articles, respectively.

Next, India, France, Spain, Japan and United Kingdom, participated each in 5.8-6.5% of the top cited papers. Finally, Italy, Turkey and the Netherlands, participated in a 4.2%, 4.1% and 3.5% of the most cited papers, respectively.

The most cited publications is the quality indicator in which the predominance of the United States is more clear, and observed in most research fields. Although the predominance has slightly reduced with time, still United States is the country publishing, with great advantage compared to the next countries, the top-cited papers (Ho and Kahn, 2014; Chuang and Ho, 2014). The presence of Chinese publications among the top cited has been demonstrated to be lower than its correspondent number of publications in many research areas (Ma et al., 2013; Chuang and Ho, 2014), however, in the Chemical Engineering area is already the 3$^{rd}$ ranked country, despite the larger number of publications were in the last few years of the analyzed period and newly published articles require time to accumulate citations.

### 3.3. International collaboration

The nature and magnitude of collaboration vary from one discipline to another, and depend upon such factors as the nature of the research problem, the research environment, and demographic factors (Subramanyam, 19893). However, it is usually found that the degree of international collaboration enhances the scientific production and produce publications of higher impact and visibility (Narin, 1991; Glänzel, 2000; Persson et al., 2004; Gazni et al., 2012). International co-authorship, in general, results in publications with higher citation rates than purely domestic papers, although this behaviour depends largely on the scientific area considered (Glänzel and Schubert, 2001; Leimu and Koricheva, 2005; Chuang and Ho, 2014; Chen and Ho, 2015). Furthermore, as it is going to be demonstrated here, international collaboration has not the same influence on publication profiles and citation impact for different countries (Ma et al., 2013).

### 3.3.1. International collaboration by regions

Fig. 8 shows the international collaboration among geographical areas during the period 2000-2011. North America collaborated with 108 countries in 11,481 publications. Its international collaboration rate (publications with international collaboration / total publications) was 28.3%. The main collaboration of North America is with Asia and Europe, with a 46.5% and 39.9%, respectively, of total North American international collaborations. Central and South American countries collaborated with 60 countries in 3,102 publications, which means an international collaboration rate of 30.6%. The main part of the collaborations were carried out with European countries (63.8%) and North America (27.0%), while the cooperation with Asia is rather limited (7.2%).

Europe, considered as a whole, collaborated with 94 countries in 13,145 articles, meaning an international collaboration rate of 17.8%. The international collaborations were distributed as follows: 39.7% with Asia, 31.6% with North America, 14.4% with Central and South America, 9.4% with Africa and 0.9% with Oceania. It is remarkable, compared to other different geographical areas, the importance of the collaborations with countries from Central and South America due to cultural and historical reasons. Furthermore, the low degree of international collaborations compared to other regions is a consequence of the large degree of international collaborations within European countries. In fact, Europe is the geographical area with more internal collaborations between countries. A 43.6% from total publications in the case of the United Kingdom are made in cooperation with other European countries, 46.1% for France, 53.3% for Germany, 56.3% for Spain, 59.9% for the Netherlands, 63.6% for Italy, 69.3% for Poland and 73.1% for Romania.

Africa cooperated with 66 countries in 1,817 publications (international collaboration rate of 40.4%). The main part of the cooperation are with Europe (58.6%) and Asia (23.7%), while the cooperation with North America represents only 12.7% of its international collaboration. Oceania is the country with the highest international collaboration rate (43.2%), publishing in international collaboration 2,463 articles. They have publications in cooperation with 64 different countries. Most of its international collaborations are with Asia (47.7%), but still important collaborations degrees with Europe (28.7%) and North America (19.2%).

Finally, Asia collaborated in 12,394 articles with 70 countries around the world. The international collaboration rate was 14.5%. The main part of the collaborations of Asia are with North America and Europe, with very similar percentages: 41.9% and 43.9%, respectively. Next, Oceania (9.0%), Africa (3.5%) and Central and South America (1.8%).

All these data suggest a very strong relation between Africa and Central and South America with Europe, and between Oceania and Asia. In the case of North America, Asia and Europe, their main international cooperations are distributed among them, with very similar shares.

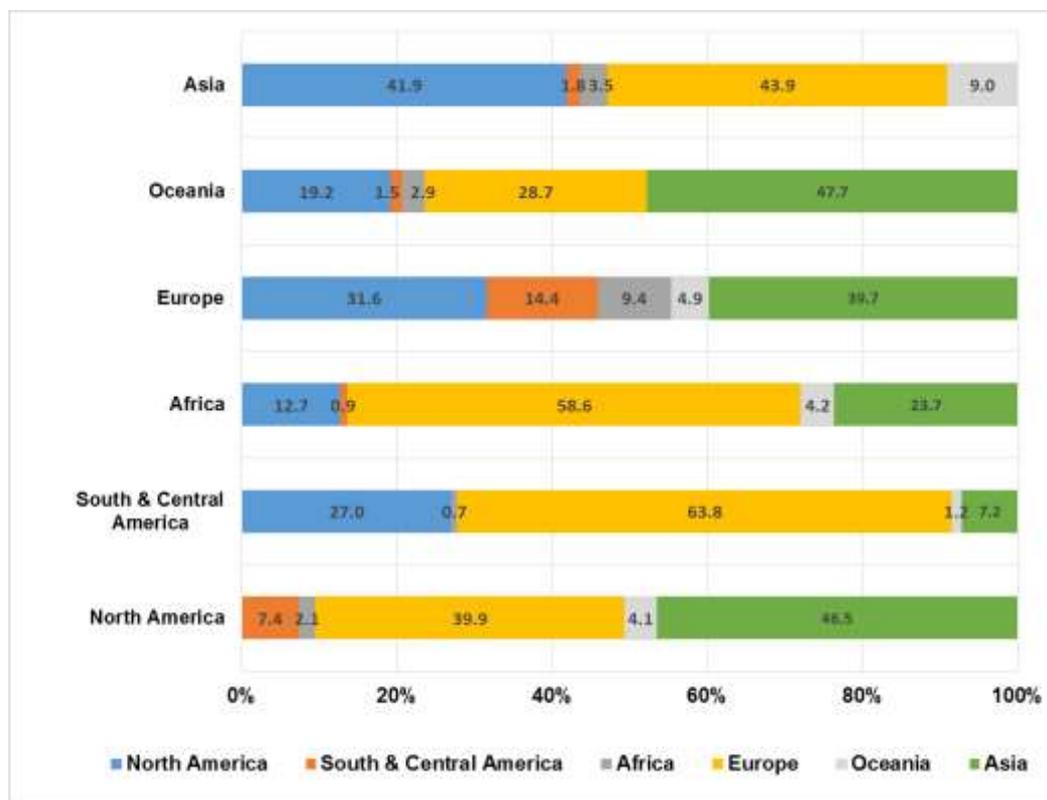

**Figure 8.** International collaboration among geographical areas during the period 2000-2011.

### 3.3.2. *International collaboration by countries*

If countries are considered separately, the number of international collaborations in publications varied widely depending on the considered country (Fig. 9). The average international collaboration among the 30 most productive countries was 34.2%. According to the international collaboration degree, the countries were classified in three main groups: a) low international collaboration (<25 per 100 papers): South Korea, Japan, Russia, Romania, Iran, China, Poland, India, Taiwan and Turkey; b) intermediate international collaborations (25-40 per 100 papers): Brazil, Egypt, United States, Argentina, Greece, Italy, Germany, Mexico, Spain and Canada; and c) high international collaborations (>40 per 100 papers): Belgium, Switzerland, the Netherlands, France, Sweden, Portugal, Singapore, United Kingdom, Finland and Australia.

According to the literature, the degree of international collaboration in Chemical Engineering is relatively high compared to other areas, however, comparisons must take into account that the degree of international collaborative publications have grown with time. Publications focused on older data will show lower international collaboration than those focused on more recent data, i.e. the international collaborative degree in the area of volatile organic compounds (VOCs) increased from around 8% in 1992 to 20% in 2007 (Zhang et al., 2009); from 8% in 1991 to around 17% in 2010 in the area of oxidative stress research (Wen and Huang, 2011); or from 9% in 1991-94 to 26% in 2003-06 in aerosol research area (Xie et al., 2008). This ascending trend of collaborative articles proportion to world publication was explained as somewhat owing to the rising number of institutes and countries engaged the research (Xie et al., 2008).

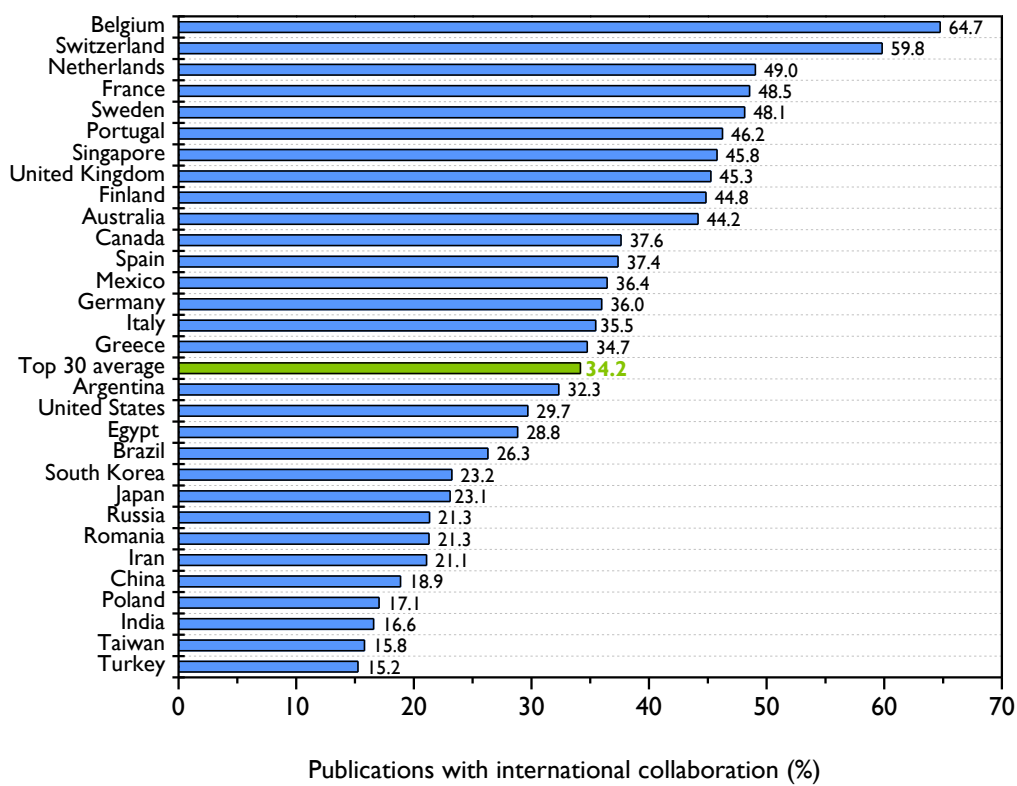

**Figure 9.** Publications made in internal collaboration by countries during the period 2000-2011.

The average international collaboration in Chemical Engineering (34.2%) was significantly higher than: 24% in proteomics (1995-2010) (Tan et al., 2014), 20% in eutrophication research (1991-2010) (Yi and Jie, 2011), 17% in oxidative stress research (1991-2010) (Wen and Huang, 2011), 16% in the area of desalination research (from 1900 to 2008) (Tanaka and Ho, 2011), 14% in biosorption technology for water treatment (1991-2004) (Ho, 2008) or the 12% in stem-cell research (1991-2006) (Li et al., 2009).

It is often argued that the international collaboration is highest within the nearest countries. In this sense, most of the countries with the highest international collaboration are European countries, where more countries are closer in the territory and where the European Union research and development programmes have efficiently promote the collaboration within European countries (Wagner and Leydesdorff, 2005; Almeida and Pais, 2009; Zitt et al., 2012;). In these countries, the intra-European collaboration (EU-27) represented the highest share of international collaborations.

Due to its importance, the international collaborations of the United States and China, the two most productive countries in the analyzed period, have been analyzed separately. The ten countries with a highest collaboration with the United States are: China (1,113, 11.5%), South Korea (729, 7.5%), United Kingdom (686, 7.1%), Germany (637, 6.6%), Canada (626, 6.5%), Japan (467, 4.8%), France (378, 3.9%), Spain (311, 3.2%), Australia (304, 3.1%) and India (300, 3.1%). They represented together a 57.4% of total international collaborations of the United States. In the case of China, the collaborations are more focused in a few number of countries, in fact, ten top countries in terms of collaboration with China represent 84.7% of total collaborations. The ten countries with a highest collaboration with China are: United States (1,113 publications, 22.6%), Japan (623, 12.7%), Canada (433, 8.8%), United Kingdom (414, 8.4%), Australia (366, 7.4%), Singapore (306, 6.2%), Germany (272, 5.5%), South Korea (233, 4.7%), France (217, 4.4%) and Taiwan (192, 3.9%).

A large degree of collaboration between the two most productive countries, United States and China, was found (1,113 publications). However, while United States represented for China around 22.6% of total international collaborations, China only represented 11.5% of the total collaborations of the United States. Whatever the case, the increase of China-United States cooperation has increased largely the last years in all the areas. If in 1986-1997, China-United States articles made up 2.5% of internationally co-authored papers from the United States, in 2008 this value increased up to 10.4% (China was the second country in the international collaborations of the United States in this year) and by 2011, China become the top collaborating country with United States (Wagner et al., 2015).

Finally, Fig. 10 shows the collaboration network of the 30 most productive countries generated by UCINET software. The nodes represent the different countries while the links represent the collaboration between countries. The size of each node indicates the total number of publications in the period while the links width indicates the number of publications in collaboration.

### 3.4. Analysis of the journals of the area

The number of journals covered by JCR in the Chemical Engineering subject category varied in the period 2000-2011, reaching a value of around 130 journals in the latest years. Table 2 shows the top 25 journals with higher number of published articles and reviews. Although there were identified around 190 different journals in the analyzed period, top 10 journals covered a 29.7% publications, top 25 journals a 50.9% and the top 50 journals, a 72.2%.

As Table 2 shows, the journal with the highest number of publications is *Industrial & Engineering Chemistry Research*, with 5.80% of total papers published in the area, followed by *Desalination*, *Chemical Engineering Science* and the *Journal of Membrane Science*, representing each around 3.0-3.3% of total papers published in the area.

According to Bradford's Law of Scattering, if the journals are sorted in descending order in terms of number of articles and then divided in three groups (each group representing one third of the total number of articles), the number of journals in each zone should follow the ratio $1:n:n^2$. This law has demonstrated to be valid previously in some research areas such as desalination (Tanaka and Ho, 2011) or proteomics (Tan et al., 2014). In the present study, the 12 first journals published 71,737 publications (33.64%), the following 30 journals published 71,292 articles (33.4%) and finally, the last 148 journals published the other 72,290 articles (33.9%). According to that, the ratio 1:2.5:12.33 ($1:n:2.0n^2$) was obtained, which is different to the theoretical Braford's Law ($1:n:n^2$).

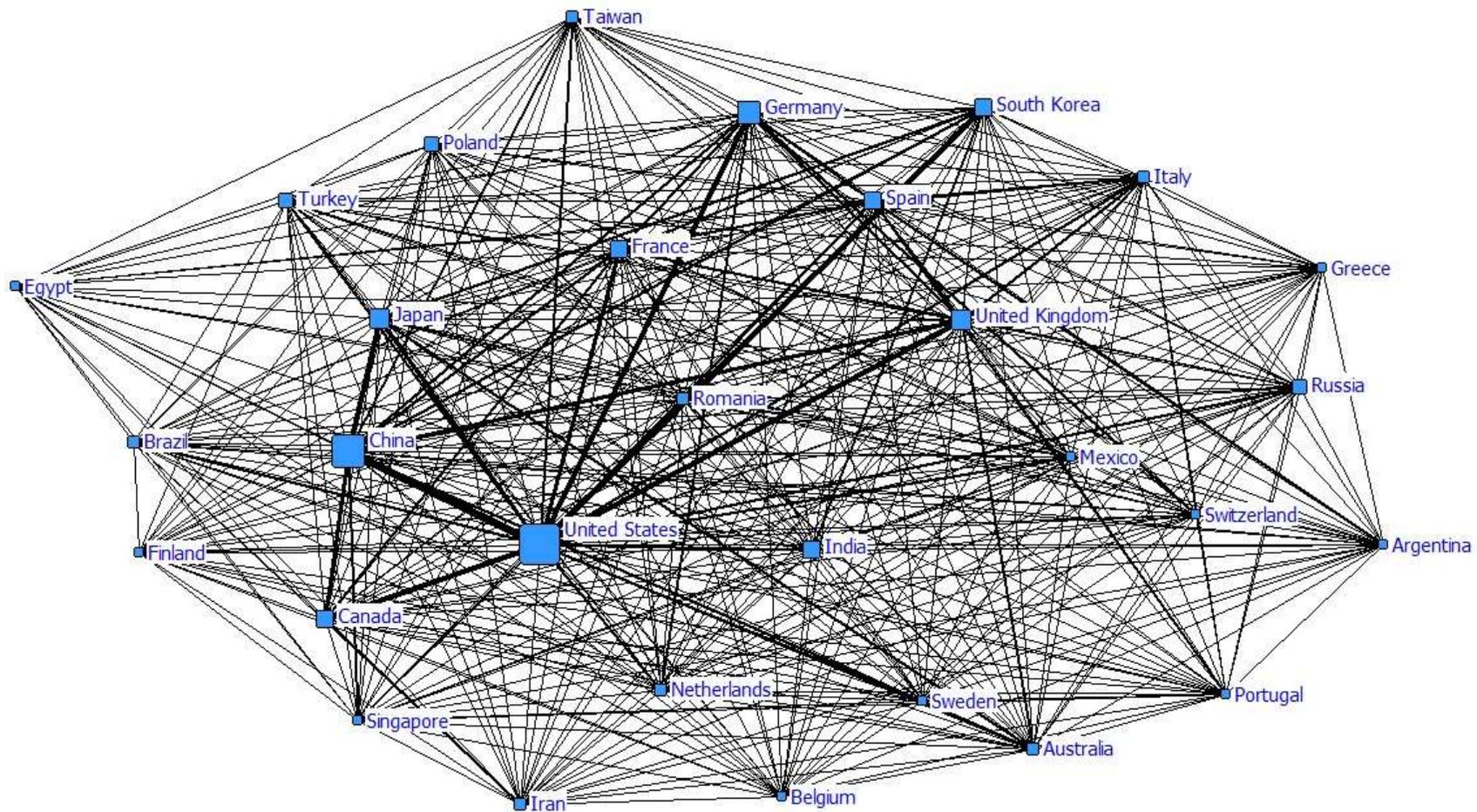

**Figure 10.** Collaboration network among the 30 most productive countries during the period 2000-2011. The size of the nodes is proportional to its total publications in 2000-2011 while the width of the ties is proportional to their number of collaborations.

Table 2. Top 25 journals in Chemical Engineering in terms of publications during the period 2000-2011. Notes: P2000-2011 = Total publications in 2000-2011; % TP = Percentage of total publications.

| Rank | Journal | PUB 2000-2011 | % TP | Publisher | Journal country | Areas | Average IF 2000-2011 | Median IF 2000-2011 |
|---|---|---|---|---|---|---|---|---|
| 1 | Industrial & Engineering Chemistry Research | 12477 | 5.80 | American Chemical Society | United States | Engineering, Chemical | 1.614 | 1.511 |
| 2 | Desalination | 7030 | 3.27 | Elsevier Science BV | Netherlands | Engineering, Chemical Water Resources | 1.152 | 0.936 |
| 3 | Chemical Engineering Science | 6862 | 3.19 | Pergamon-Elsevier Science Ltd. | United States | Engineering, Chemical | 1.794 | 1.770 |
| 4 | Journal of Membrane Science | 6561 | 3.05 | Elservier Science BV | Netherlands | Engineering, Chemical Polymer Science | 2.662 | 2.543 |
| 5 | Chemical Engineering News | 5580 | 2.60 | American Chemical Society | United States | Chemistry, Multidisciplinary Engineering, Chemical | 0.413 | 0.381 |
| 6 | Journal of Chemical and Engineering Data | 5526 | 2.57 | American Chemical Society | United States | Chemistry, Multidisciplinary Engineering, Chemical | 1.479 | 1.626 |
| 7 | Catalysis Today | 5400 | 2.51 | Elsevier Science BV | Netherlands | Chemistry, Applied Chemistry, Physical Engineering, Chemical | 2.696 | 2.696 |
| 8 | Energy Fuels | 5189 | 2.41 | American Chemical Scoiety | United States | Energy & Fuels Engineering, Chemical | 1.706 | 1.507 |
| 9 | Chemical Engineering Journal | 4902 | 2.28 | Elsevier Science SA | Switzerland | Engineering, Environmental Engineering, Chemical | 1.839 | 1.651 |
| 10 | Journal of Food Engineering | 4330 | 2.01 | Elsevier Sci Ltd. | United Kingdom | Engineering, Chemical Food Science & Technology | 1.563 | 1.473 |
| 11 | Journal of Catalysis | 4005 | 1.86 | Academic Press Inc. Elsevier Science | United States | Chemistry, Physical Engineering, Chemical | 4.516 | 4.737 |
| 12 | Fuel | 3875 | 1.80 | Elsevier Sci. Ltd. | United Kingdom | Energy & Fuels Engineering, Chemical | 1.897 | 1.521 |
| 13 | Fluid Phase Equilibria | 3470 | 1.61 | Elsevier Science BV | Netherlands | Thermodynamics Chemistry, Physical Engineering, Chemical | 1.519 | 1.492 |
| 14 | Powder Technology | 3434 | 1.60 | Elsevier Science SA | Switzerland | Engineering, Chemical | 1.271 | 1.175 |
| 15 | AIChE Journal | 3394 | 1.58 | Wiley-Blackwell | United States | Engineering, Chemical | 1.868 | 1.838 |
| 16 | Applied Catalysis B Environmental | 3382 | 1.57 | Elsevier Science BV | Netherlands | Chemistry, Physical Engineering, Environmental Engineering, Chemical | 4.161 | 3.992 |
| 17 | Process Biochemistry | 3060 | 1.42 | Elsevier Sci. Ltd. | United Kingdom | Biochemistry & Molecular Biology, Biotechnology & Applied Microbiology Engineering, Chemical | 1.792 | 1.902 |
| 18 | Separation and Purification Technology | 2876 | 1.34 | Elsevier Science BV | Netherlands | Engineering, Chemical | 2.259 | 2.374 |
| 19 | Polymer Engineering and Science | 2849 | 1.33 | Wiley-Blackwell | United States | Engineering, Chemical Polymer Science | 1.131 | 1.235 |
| 20 | Revista De Chimie | 2849 | 1.33 | Chiminform Data S.A. | Romania | Chemistry, Multidisciplinary Engineering, Chemical | 0.358 | 0.289 |
| 21 | Separation Science and Technology | 2681 | 1.25 | Taylor & Francis Inc. | United States | Chemistry, Multidisciplinary Engineering, Chemical | 0.927 | 0.893 |
| 22 | Korean Journal of Chemical Engineering | 2538 | 1.18 | Korean Institute of Chemical Engineers | South Korea | Chemistry, Multidisciplinary Engineering, Chemical | 0.784 | 0.813 |
| 23 | Chemical Engineering Technology | 2413 | 1.12 | Wiley – VCH Verlag GmbH | Germany | Engineering, Chemical | 0.884 | 0.857 |
| 24 | Chemie Ingenieur Technik | 2367 | 1.10 | Wiley – VCH Verlag GmbH | Germany | Engineering, Chemical | 0.338 | 0.368 |
| 25 | Journal of Chemical Technology and Biotechnology | 2309 | 1.07 | Wiley-Blackwell | United Kingdom | Biotechnology & Applied Microbiology Chemistry, Multidisciplinary Engineering, Chemical | 1.342 | 1.129 |

Regarding the subject categories in which the different journals have been indexed in WoS, it is observed that most of the journals are included in several subject categories. In fact, only five of the 25 journals with the highest number of publications are indexed only in Chemical Engineering subject category (25%). However, it is remarkable that two of the journals indexed only in Chemical Engineering subject category are ranked in the first positions of the list: *Industrial & Engineering Chemistry Research* (rank 1$^{st}$) and *Chemical Engineering Science* (ranked 3$^{rd}$).

Other interesting aspect to analyze are the publishers and their nationalities. The publishers of the top 25 journals in Chemical Engineering are Elsevier (11), the American Chemical Society (4), Wiley-Blackwell (3) and Wiley VCH-Verlag (2). The rest of publishers edited only one journal each: Pergamon-Elsevier, Academic Press-Elsevier, Chiminform Data, Taylor & Francis and the Korean Institute of Chemical Engineers. Regarding the journal countries, there are seven different nationalities, the most important being United States (9 journals), the Netherlands (6) and United Kingdom (4). The rest of the countries are Switzerland (2), Germany (2), Romania (1) and South Korea (1).

If the 133 journals covered by JCR in 2011 are considered, there a total of 58 different publishers were identified. The top publishers were: Elsevier (31 journals), Wiley-Blackwell (13), Taylor & Francis Ltd. (12), Springer (7), Wiley-VCH Verlag GmbH (5), American Chemical Society (4) and Pergamon-Elsevier Science Ltd. (4). In this case, a total of 26 different nationalities of the journals were obtained, but very similar distribution than in the case of the top 25 journals was obtained. United States is the first publisher (38 publications, 28.6%), followed by United Kingdom (27, 20.3%), the Netherlands (14, 10.5%), Germany (10, 7.5%), Japan (6, 4.5%), China (5, 3.8%), Russia (5, 3.8%), Switzerland (4, 3.0%), and Poland (3, 2.3%).

It was also observed that there are some countries, which publish one or two journals, in which the national production is highly concentrated. For example, an 80.8% of scientific production from Romania is published in the only Romanian journal covered in JCR (*Revista De Chimie*) and something similar occurs in Uruguay, where a 58.2% of scientific production is published in the only Uruguayan journal covered in JCR (*Ingeniería Química*). The situation in Macedonia, Serbia, etc. is also very similar to that of Romania or Uruguay. Furthermore it is still interesting to notice that around 58% of Russian papers were published in the five Russian journals covered in JCR, the 48% of papers from Poland were published in their 3 Polish journals and the 25% of Chinese papers were published in the six Chinese journals covered in the JCR. As recognized by Fu et al. (2014), despite the great progress in the Chinese research in Chemical Engineering, the growth of China's publications is also partially due to the increasing number of Chinese journals recorded by international databases recently.

### 3.5. Research organizations

Due to the lack of standardization in the organizations names, the analysis of organizations is always complex, especially taking into account the large number of documents to be analyzed. For this reason, WoS offers the possibility to search not only by "organization" but also by "organization-enhanced" since later versions than 5.9. This "organization-enhanced" search allow preferred organization names to be searched in order to deliver a more comprehensive set of results by searching variant organization names.

Different results were obtained when using "organization" and "organization-enhanced" searches. Although the results obtained are similar in many cases, it is interesting to notice that the organization with the highest number of publications, for example, is different depending on the used search. While the *Chinese Academy of Sciences* is the most productive institution if "organizations" search is used (3,787 publications), the *Centre National de la Recherche Scientifique* (CNRS) is the most productive institution if "organizations-enhanced" is used (4,518 publications), while it was ranked 12$^{th}$ ranked using "organizations" search (986 publications). Another important difference is related to the *United States Department of Energy* (DoE), ranked 4$^{th}$ in "organizations-enhanced" search (2,527 publications) while did not appeared among the 100 most productive organizations in "organizations" search. The main differences observed using these two types of search are in large organizations, with many variations in the name of the organizations, where the "organizations-enhanced" search allows a more accurate analysis. This is the reason why this search is the one selected in this paper to analyze in detail the organizations involved in Chemical Engineering R&D.

Table 3 shows the 40 institutions with the highest number of publications (>725 publications) along the period 2000-2011. There are organizations from 21 different countries among and these organizations participated in 54,750 publications, representing a 25.7% of total publications. Among the 40 institutions with the highest number of publications, the following distribution was obtained: China (8), United States (5), France (4), India (2), Germany (2), Japan (2), South Korea (2), United Kingdom (2), Russia (1), Spain (1), Singapore (1), Netherlands (1), Argentina (1), Romania (1), Taiwan (1), Brazil (1), Canada (1), Switzerland (1), Portugal (1), Denmark (1), Italy (1). The top five institutions are from France, China, India, United States and Russia.

**Table 3**.- Top 40 organizations in terms of number of publications in the Chemical Engineering area during 2000-2011.

| Rank | Organization | Country | PUB | % PUB |
|---|---|---|---|---|
| 1 | Centre National de la Recherche Scientifique (CNRS) | France | 4,518 | 2.12 |
| 2 | Chinese Academy of Sciences | China | 3,884 | 1.82 |
| 3 | Indian Institutes of Technology (all 7 IITs together) | India | 2,678 | 1.26 |
| 4 | United States Department of Energy DOE | United States | 2,527 | 1.18 |
| 5 | Russian Academy of Sciences | Russia | 2,472 | 1.16 |
| 6 | Council of Scientific Industrial Research (CSIR) | India | 2,385 | 1.12 |
| 7 | Tsing Hua University | China | 1,890 | 0.89 |
| 8 | University of California System | United States | 1,764 | 0.83 |
| 9 | Consejo Superior de Investigaciones Científicas (CSIC) | Spain | 1,711 | 0.80 |
| 10 | Zhejiang University | China | 1,608 | 0.75 |
| 11 | Tianjin University | China | 1,504 | 0.71 |
| 12 | National University of Singapore (NUS) | Singapore | 1,401 | 0.66 |
| 13 | East China University of Science Technology (ECUST) | China | 1,302 | 0.61 |
| 14 | Delft University of Technology | Netherlands | 1,230 | 0.58 |
| 15 | National Institute of Advanced Industrial Science Technology (AIST) | Japan | 1,222 | 0.57 |
| 16 | Consejo Nacional de Investigaciones Científicas y Técnicas (CONICET) | Argentina | 1,179 | 0.55 |
| 17 | Imperial College London | United Kingdom | 1,137 | 0.53 |
| 18 | Pennsylvania Commonwealth System of Higher Education (PCSHE) | United States | 1,051 | 0.49 |
| 19 | Pres Universite de Toulouse | France | 1,048 | 0.49 |
| 20 | Polytechnic University of Bucharest | Romania | 1,045 | 0.49 |
| 21 | Beijing University of Chemical Technology | China | 960 | 0.45 |
| 22 | Dalian University of Technology | China | 956 | 0.45 |
| 23 | Seoul National University | South Korea | 942 | 0.44 |
| 24 | University of Lorraine | France | 941 | 0.44 |
| 25 | National Taiwan University | Taiwan | 920 | 0.43 |
| 26 | University of Alberta | Canada | 915 | 0.43 |
| 27 | Universidade Estadual de Campinas | Brazil | 911 | 0.43 |

| 28 | China University of Petroleum | China | 902 | 0.42 |
|---|---|---|---|---|
| 29 | Karlsruhe Institute of Technology | Germany | 876 | 0.41 |
| 30 | University of Leeds | United Kingdom | 865 | 0.41 |
| 31 | Swiss Federal Institute of Technology Zurich (ETH Zurich) | Switzerland | 857 | 0.40 |
| 32 | Korea Institute of Technology | South Korea | 838 | 0.39 |
| 33 | Penn State University | United States | 838 | 0.39 |
| 34 | Universidade do Porto | Portugal | 830 | 0.39 |
| 35 | Technical University of Denmark (DTU) | Denmark | 827 | 0.39 |
| 36 | Consiglio Nazionale delle Ricerche (CNR) | Italy | 818 | 0.38 |
| 37 | IFP Energies Nouvelles | France | 762 | 0.36 |
| 38 | Max Planck Society | Germany | 756 | 0.35 |
| 39 | University System of Georgia | United States | 752 | 0.35 |
| 40 | Toohoku University | Japan | 728 | 0.34 |

China has eight institutions in the 40 most productive institutions, three of them among the 10 most productive (*Chinese Academy of Sciences*, *Tsing Hua University* and *Zhejian University*). United States has five institutions in the 40 most productive institutions, 2 of them among the 10 most productive (*United States Department of Energy* and *University of California System*). France has four institutions among the 40 most productive institutions, including the first ranked. India has two institutions among the 40 most productive institutions but both are in the top 10 institutions. They are the *Indian Institutes of Technology* (IIT) and the *Council of Scientific Industrial Research* (CSIR). Apart from China, India and United States, Spain and Russia completes the list of countries with institutions among the top 10. Russia has the *Russian Academy of Sciences* ranked 5[th], and Spain, the *Consejo Superior de Investigaciones Científicas* (CSIC) ranked 9[th]. Finally, Germany, Japan, South Korea or United Kingdom have two institutions among the 40 most productive institutions, but none among the top 10 institutions. The rest of the countries among the 40 most productive institutions with only one institution.

In a previous study carried out by Modak and Madras (2008), they found that the top 10 institutions in Chemical Engineering in 2006 were from China (5), United States (4), and Taiwan (1). In the study of Modak and Madras, it was observed that the Chinese organizations increased largely its importance in the period 1990-2006 while the importance of organizations from the United States decreased. For example, the 1[st] ranked organization in 2006 was *Tsing Hua University* which was ranked the 425[th] in 1990-1994. Similarly, *Tianjin University* was ranked the 47[th] in 1990-1994 but 2[nd] in 2006, and the *Chinese Academy of Sciences* was ranked 320[th] 1990-1994 but 3[rd] in 2006. On the other hand *MIT*, the 2$^{nd}$ institution in 1990-94, was ranked 6$^{th}$ in 2006; *University of Texas* decreased its importance from the 4$^{th}$ ranked in 1990-1994 to the 7$^{th}$ in 2006 and the *California Institute of Technology* decreased its rank from the 3$^{rd}$ in 1990-94 to the 9$^{th}$ in 2006. The results of Modak and Madras (2008) are not directly comparable with the results obtained in the present study as the data they used were obtained before the "organizations-enhanced" tool was available at WoS, therefore, large institutions as the CNRS or the United States Department of Energy were underestimated in their study.

If the analysis is extended to the top 100 institutions, a total of 31 different nationalities are found, these institutions participating in 91,667 publications (43.0% total publications). By countries, the number of institutions among the top 100 are the following: China (12), United States (12), France (9), Japan (7), United Kingdom (7), Canada (5), Australia (5), Germany (4), Poland (4), South Korea (4), India (3), Netherlands (3), Brazil (3), Spain (2), Singapore (2), Portugal (2), Mexico (2), Russia (1), Taiwan (1), Switzerland (1), Denmark (1), Italy (1), Thailand (1), Turkey (1), Greece (1), Norway (1), Iran (1), Sweden (1), Romania (1), Czech Republic (1) and Argentina (1).

Fig. 11 shows the relationship between the number of publications by countries and the number of organizations of these countries among the top 100 institutions. Countries located above the fit of the date (shown in dashed line) are countries with a high number of small organizations involved in R&D in Chemical Engineering while those below the fit are countries with the scientific production concentrated in only a few institutions.

On one side, India, Spain and Germany have intermediate-high scientific production but a relative low number of institutions among the top 100. India has 10,145 publications (6$^{th}$ ranked in publications), but with only three organizations among the top 100. The most active institutions of India in Chemical Engineering are very large especially the Indian Institutes of Technology (ranked 3$^{rd}$) and the Council of Scientific Industrial Research (ranked 6$^{th}$). The situation for Germany and Spain is similar, they have intermediate-high scientific production but a low number of institutions among the top 100, four in the case of Germany and two in the case of Spain (including the CSIC, ranked 9$^{th}$), meaning the R&D is concentrated in a small number of organizations.

On the contrary, France has nine institutions among the top 100 despite its scientific production is intermediate, meaning the publications are diversified in a greater number of R&D organizations. Although the French CNRS is involved in around 45% of the total publications in this area, there are eight more French organizations among the top 100 most productive in the area. Similar behavior to that of France was observed for Canada and Australia.

If the most productive countries, United States and China, are compared again, it was found that both have the same number of research organizations among the top 100 (12) while United States had a 25% greater scientific production than China in the period 2000-2011. This means the scientific production of China is most concentrated in some research organizations than the United States. This is in agreement with the fact that there are five Chinese institutions and only two institutions from the United States among the top 15, despite they have the same number of institutions among the top 100.

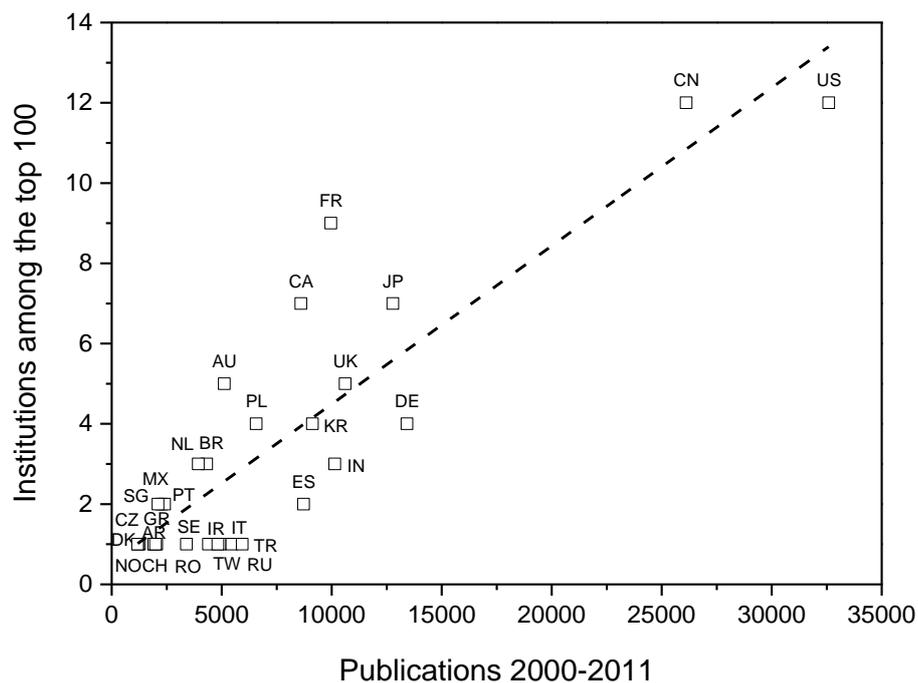

**Figure 11.** Number of institutions among the top 100 vs. number of publications in the period 2000-2011 by countries. Note: Countries are named according two letter codes of ISO 3166-1.

## 4. CONCLUSIONS

The main conclusions of this study are the following:

- The contribution of Asia, Africa, Central and South America and Oceania to the world scientific production have increased largely in the period 2000-2011, while the contribution of Europe, and especially North America, have decreased.
- An important displacement of the scientific production to the Far East has been observed in the analyzed period, due to the increase of publications of China but also for countries such as India and Iran.
- Countries such as Switzerland, Italy, Netherlands, Singapore and Spain have considerably higher impact factor and number of cites per article and year since published than the world average and the 30 most productive countries. The opposite is true for Romania, Russia, Poland, Egypt and Iran.
- United States outstands as the country with the highest number of articles among the 1000 most cited (31.5%), followed by Germany (8.4%) and China (7.5%).
- The international collaboration among continents and countries is considerably important in the area. European countries are those with a higher international collaboration rate, outstanding Belgium, Switzerland, Netherlands, France, Sweden and Portugal with international collaboration rates higher than 50%.
- The scientific production is concentrated in a few journals, top 10 journals publishing around 30% publications and top 25 journals, around 50%. Most of the journals (around 75%) are indexed in several subject categories in WoS however, 1$^{st}$ and 3$^{rd}$ ranked journals in terms of publications are indexed only in the Chemical Engineering subject category. Elsevier, Wiley-Blackwell and Taylor & Francis are the most important publishers. The American Chemical Society is also a very important publisher, it publishes only four journals but all these four are among the top 10 in terms of number of publications.
- The five most productive institutions are from France (*CNRS*), China (*Chinese Academy of Sciences*), India (*Indian Institutes of Technology*), United States (*Department of Energy*) and Russia (*Russian Academy of Sciences*). If the top 100 institutions are considered, China and United States are those with a higher number (12 each), followed by France (9), Japan (7) and United Kingdom (7).